# LOCALIZATION FOR WIRELESS SENSOR NETWORKS: A NEURAL NETWORK APPROACH

Shiu Kumar[1,*], Ronesh Sharma[2] and Edwin R. Vans[3]

[1,2,3] Department of Electronics Engineering, Fiji national University, Suva, Fiji
[1,2] School of Engineering & Physics, University of the South Pacific, Suva, Fiji

## ABSTRACT

*As Wireless Sensor Networks are penetrating into the industrial domain, many research opportunities are emerging. One such essential and challenging application is that of node localization. A feed-forward neural network based methodology is adopted in this paper. The Received Signal Strength Indicator (RSSI) values of the anchor node beacons are used. The number of anchor nodes and their configurations has an impact on the accuracy of the localization system, which is also addressed in this paper. Five different training algorithms are evaluated to find the training algorithm that gives the best result. The multi-layer Perceptron (MLP) neural network model was trained using Matlab. In order to evaluate the performance of the proposed method in real time, the model obtained was then implemented on the Arduino microcontroller. With four anchor nodes, an average 2D localization error of 0.2953 m has been achieved with a 12-12-2 neural network structure. The proposed method can also be implemented on any other embedded microcontroller system.*

## KEYWORDS

*Levenberg-Marquardt (LM) Algorithm, Localization, Neural Network, Received Signal Strength Indicator (RSSI), Wireless Sensor Network (WSN).*

## 1. INTRODUCTION

Wireless Sensor Networks (WSNs) has gained vast response from academia and industries with new applications being developing in numerous fields. Some of the prospective applications include environmental and vegetation monitoring, search and rescue operations [12], object tracking such as tracking patients and doctors in hospitals, monitoring patients, military applications [10] and other industrial applications. As a result of advancement in technologies, sensor nodes that are small in size, are cheaper and consume less power having capabilities such as sensing, data storage, computing and wireless communication have been developed.

An essential and challenging part of many WSN applications is localization (process of finding the position of the nodes [8]) amongst the others such as architecture, deployment, synchronization, calibration, quality of service and security. The sensor data has to be attached with the measured data to make it significant for localization as this is essentially required in monitoring and recording a wide-ranging information such as acoustic, thermal, visual, seismic or any other type of measured observation. For example, a person monitoring a large vegetation farm with a huge variety of vegetables would require the location or the part of the farm from which the data is received as different vegetables have different specific requirements. To report the origin of objects, node localization is necessary so that group querying of sensors, routing, and questions about the network coverage [1] can be assisted or answered. Over the decade, a number of solutions have been provided by researches for the problem of node localization (for example, [2]-[6], [9], [11], [13]-[16], [18]-[21], [23], [25]-[26], [33]-[34], [36]). In general, mostly there is a trade-off between the accuracy of localization, the computational complexity, and energy efficiency (for example, [24], [28-32]) of the techniques depending on their application requirements.

A WSN usually comprises of a huge number of spatially distributed sensor nodes making it difficult and impractical to record the sensor node locations when the nodes are being deployed. There is also a possibility that the sensor node might later be moved from its original location to some other location. Therefore, an algorithm that will autonomously determine the location of the nodes is required. A novel neural network based node localization scheme is proposed in this research work. The Received Signal Strength Indicator (RSSI) has been used for estimating the coordinates or position of wireless sensor nodes in order to tackle the problem of determining the location of the node. The following contributions are made:

- Is RSSI a suitable component to be used for node localization in WSNs? We employ the use of RSSI value to determine the location of the sensor nodes and explain whether RSSI is suitable to be used for localization purposes (Section 4). The need for additional hardware will not arise as sensor nodes are freely equipped with RF modules for wireless communication; therefore no extra hardware will be required. However, multipath fading and noise will contaminate the RF signals, thus affecting the RSSI value by corrupting the signal to noise ratio.
- The choice of the evaluation model is also of paramount importance. In this research, a 2D indoor environment is considered, that is the location is determined by the *x* and *y* coordinates. A single anchor node will require steerable directional antennas, thus accounting for a direct increase in cost and power consumption of the sensor node. Therefore, two or more anchor nodes with different configurations (Section 3.2) have been employed and the best configuration to be used for the anchor nodes is explained (Section 4).
- We introduce a novel neural network based 2D indoor localization scheme. Global Positioning System (GPS: [4], [19]) can be used for accurate localization. However, use of GPS requires an external hardware to be attached to the sensor nodes leading to an increase in cost and power consumption. Moreover, line-of-sight is required for GPS and thus is not applicable for indoor environments. The method of indoor localization proposed uses a feed-forward neural network (See Section 2). A number of neural network structures trained using Levenberg-Marquardt (LM), Bayesian Regularization (BR), Resilient Back-propagation (RP), Scaled Conjugate Gradient (SCG) and Gradient Descent (GD), have been evaluated and the best neural network employed (refer to Section 4).

## 1.1. PROBLEM DEFINITION

Consider the case when N number of sensor nodes is deployed in a sensor network with locations L = {L$_1$, L$_2$, ..., L$_N$}. Let L$_{xi}$, L$_{yi}$, L$_{zi}$, denote the *x*, *y* and *z* coordinates of the $i^{th}$ sensor node respectively. The problem of node localization involves determining these locations and by letting L$_{zi}$ = 0, problem of 2 dimension is obtained. Sensor nodes having knowledge of their positions are known as anchors or beacons. All nodes in the network with unknown position localize themselves with the aid of these anchor nodes. Therefore, mathematically the node localization problem is stated as follows: for a given a multi-hop WSN, represented by a graph G = (V, E) with A anchor nodes having positions {$x_a$, $y_a$} for all a ε A, we want to determine the position {$x_u$, $y_u$} for all nodes with unknown positions u ε U.

## 2. THE PROPOSED NEURAL NETWORK METHOD

A practical method of learning discrete valued and real valued functions form given examples is provided by artificial neural network (ANN, also referred to as neural network). Supervised learning is used by ANN. The inputs and the outputs are provided to the ANN in order for it to learn and form a suitable model. It is generally used to tackle the problem of classification and regression. Usually, a multilayer neural network has three layers of interconnected "artificial neurons"; the input layer, hidden layer(s) and the output layer (see figure 1).

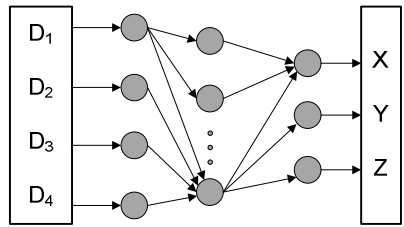

Figure 1. A general feed-forward neural network structure with 4 inputs and 3 outputs.

Feed-forward neural networks and feedback neural networks are the two groups of neural network. In a feed-forward neural network, the outputs from one layer of neurons is fed to the next layer and in this process no layers are skipped and there is no feedback to the system. The three main types of neural networks used in localization [3] are MLP, Recurrent Neural Network (RNN) and Radial Basic Function (RBF).

It must be noted that the RSSI values obtained are highly unstable and turn to vary under environmental noise and mobility of sensor nodes. A neural network offers the advantage that prior knowledge of the environment and noise distribution is not necessary. Moreover, higher accuracies are achieved by neural networks compared to other techniques such as the Kalman filter [3]. The trade-off between the accuracy and memory requirements of the MLP neural network is the best when compared with other types of neural networks, thus it has been chosen to be used in this research.

Matlab has been used for the implementation of the MLP neural network (a feed-forward ANN) with three layers. The best solution chosen (refer to Section 4) is shown in figure 2 comprising of three inputs, twelve nodes in the first and second (hidden) layers, and two nodes in the output layer. RSSI values acquired from the 3 anchor nodes are fed as the input to the system, while the output generates the estimate of the *x* and *y* coordinates of the mobile node.

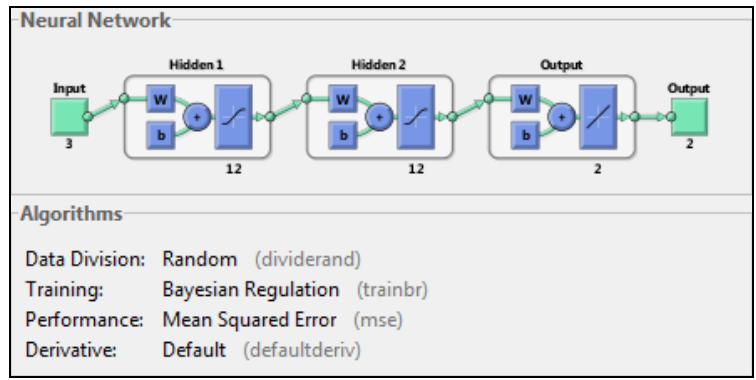

Figure 2. The proposed MPL neural network with 3 inputs and 2 outputs.

The nodes in the first and second layer use the hyperbolic tangent sigmoid ("tansig") activation function, while the third layer uses a linear ("purelin") activation function. Bayesian Regularization (BR) algorithm was used for training of the network since optimal results were obtained using this training algorithm (refer Section 4). The network training time of the BR algorithm was high; however it was still employed as only offline training was required. The structure of the matrix containing the RSSI values and the output coordinates is as follows:

$$data = \begin{bmatrix} R_{11} & R_{1j} & \cdots & R_{1n} & X_1 & Y_1 \\ R_{i1} & R_{ij} & \cdots & R_{in} & X_i & Y_i \\ \vdots & \vdots & \ddots & \vdots & \vdots & \vdots \\ R_{m1} & R_{mj} & \cdots & R_{mn} & X_m & Y_m \end{bmatrix} \tag{1}$$

Where $R_{ij}$ signifies the RSSI values of the signal obtained from the $j^{th}$ anchor, at the $i^{th}$ reference point while $X_i$ and $Y_i$ signify the $x$ and $y$ coordinates of the $i^{th}$ reference point. For this research, the total number of reference points ($m$) is 94 and the number of anchors ($n$) used is three. The neural network obtained is represented by the equation as follows:

$$\begin{bmatrix} x \\ y \end{bmatrix} = \tanh\left(\tanh(R \bullet W_{1i} + b_{1i}) \bullet W_{2j} + b_{2j}\right) \bullet W_{3k} + b_{3k} \tag{2}$$

Where $\boldsymbol{R}$ is the input row vector of length 3, which consists of the RSSI values acquired from the 3 anchor nodes, $\boldsymbol{W_{mn}}$ is the weight vector of $n^{th}$ node at the $m^{th}$ layer, and $\boldsymbol{b_{mn}}$ is the bias vector of $n^{th}$ node at $m^{th}$ layer. Equation 3 was incorporated on a mobile node equipped with an Arduino microcontroller for node localization. However, this method can also be used to implement neural network on other programming platforms.

## 3. DATA COLLECTION

### 3.1. MEASURING THE RSSI VALUES

For experimental purposes, the wireless sensor node comprising of Arduino UNO in conjunction with the XBee series 2 module that is compliant with the IEEE 802.15.4 standard called ZigBee have been used for the mobile node. The anchor nodes only comprise of the XBee series 2 modules on its own. For the wireless communication between the nodes, the standard ZigBee communication is employed with an air data rate of 256 kbps operating at the ISM (Industrial, Scientific & Medical) 2.4 GHz frequency band. The XBee has a receiver sensitivity of -96 dBm and a communication range of 40m for indoor/urban environment.

For RSSI computation, XBee Series 2 modules special IEEE 802.15.4-2003 hardware support such as has been utilized. The XBee modules have an RSSI pin, which outputs a PWM signal to represent the RSSI value. This pin value can be read by the microcontroller and converted appropriately to reflect the RSSI value. However, for this research the mobile node with XBee in API mode and the anchor nodes in the AT mode was used. In API mode, packets are constructed to command the XBee explicitly, as opposed to just sending the data serially in the AT or the "Transparent Mode". The DB command which returns the RSSI value of the latest packet received was used.

The mobile node sends a signal to the anchor nodes requesting to send localization beacon, while the anchor nodes respond by sending a beacon to the mobile node. A single instance of measurement consists of one RSSI value for each anchor node signal received on the mobile node (see figure 3). The RSSI measurements between the mobile and the anchor nodes are measured and recorded for further analysis using the Docklight (version 1.9) software running on Windows 7, which reads the serial data from the node and writes it to a file.

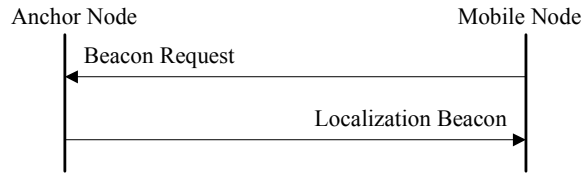

Figure 3. Message flow between mobile and anchor nodes for RSSI measurements.

## 3.2. Experimental Setup

Training and testing of a neural network requires a set of data [15], [26]. The dataset in this research is prepared by collecting the RSSI measurements for each mobile node coordinates, $r_i = [x_i, y_i]$ denoting the position of the $i^{th}$ reference point. For indoor localization, the structure of the room or experimental environment is an important factor [27]. Measurements were carried out in a research laboratory containing furniture's and equipment's such as tables, chairs and computers. The structure of the measured training data is 9 x 11 measurement points (black points) as shown in figure 4. The distance between the grid points is 0.45 m. Measurements from the training positions and from unknown positions were used for testing. The unknown positions were chosen between the training measurement points as indicated in green in figure 4.

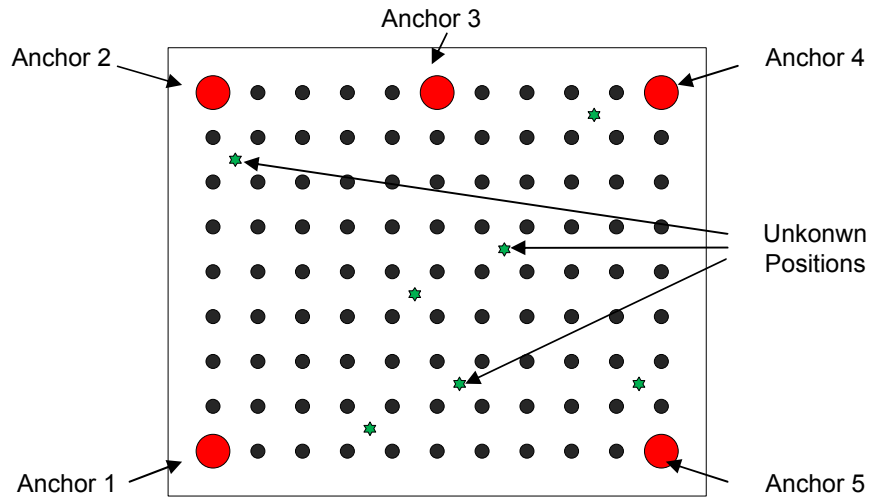

Figure 4. Experimental setup layout.

The anchor node configurations used for this research to determine the best anchor node configuration are 2 anchors (consisting of: (a) anchor 2 & 4. (b) anchor 1 & 4), 3 anchors (consisting of: (a) anchor 1, 2 & 4. (b) anchor 1, 3 & 5), 4 anchors (consisting of anchor 1, 2, 4 & 5) and 5 anchors.

## 4. RESULTS AND DISCUSSION

A neural network requires supervised learning and for this research five different training algorithms namely LM, BR, RP, SCG and GD have been used for training the neural network. A number of tests were conducted using various activation functions with varying number of layers and nodes. A network with three layers having a 12-12-2 structure gave optimal results. It employed 12 nodes in the first and second layer (which are the hidden layers) and 2 nodes in the output layer. Hyperbolic tangent sigmoid activation function has been used for the hidden layers while a pure line function has been used in the output layer. This structure was used with different number of inputs that is with different number of anchors to evaluate the performance of different anchor node configurations that follows.

An experiment was carried out to determine the best training algorithm to be used for training the neural network and was performed using 4 anchors. The data structure used for training is mentioned in Section 2 and consists of 2350 data sets of which 80 percent was used for training and validation while 20 percent was used for testing. The network was then further tested for its performance using 105 data sets obtained from 7 unknown positions (see figure 4). The performance of the network is evaluated based on the error in the distance between the estimated and exact distance. Equation 3 is used for calculating the average error.

$$e = \sum_{i=1}^{n} \frac{1}{n} \sqrt{(x_i - x_{oi})^2 + (y_i - y_{oi})^2} \tag{3}$$

Where $n$ is the total number of test sets, $(x_{oi}, y_{oi})$ is the exact position and $(x_i, y_i)$ is the estimated position of the mobile node at the $i^{th}$ test data set. Figure 5 shows the average localization error obtained in this phase. The time taken to train the network using various training algorithms is shown in figure 6.

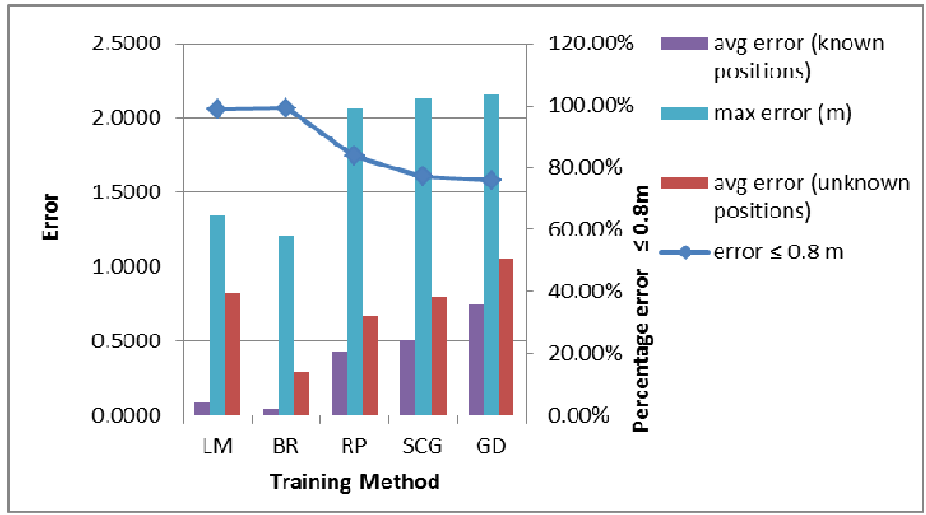

Figure 5. Localization error of different methods with 4 anchors.

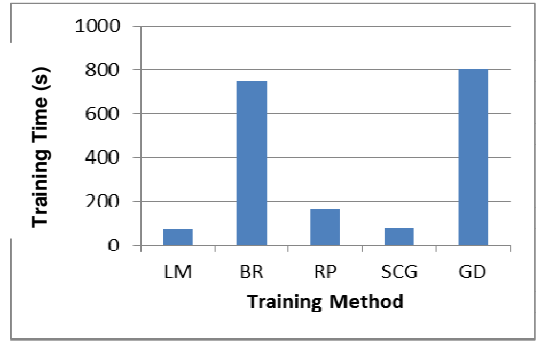

Figure 6. Time taken to train the neural network for different training algorithms with 4 anchors.

The maximum error, percentage of error less than 0.8m, average error for the test set of known positions and average error for test set of unknown positions together with the time taken to train the neural network for each method has been utilized to evaluate the performance. It can be noted that the percentage of time the error is less than 0.8 m is quite low for RP, SCG and GD training algorithms while they also have higher maximum error compared to that of LM and BR training algorithms. The training time for BR and GD training algorithms is noticeably high compared to other methods. However, for this research, offline training of the neural network is performed and the neural network is than implemented on the mobile node. Therefore, the LM method and BR method were chosen to be used for training the neural network for evaluation of other mentioned anchor node configurations.

Figure 7 shows the localization errors of the neural network for the different anchor node configurations, where the BR and LM training algorithms have been used to train the neural network. It can be noted that increasing the number of anchor nodes increases the localization accuracy. The lowest average error of all the configurations evaluated was obtained with five anchor nodes. The maximum error for neural network obtained using LM training algorithm is almost same when three, four or five anchors are used. For the two different configurations with three anchors evaluated, the second configuration with anchors one, three and five produced a slightly better result compared to that of when anchors one, two and four are used.

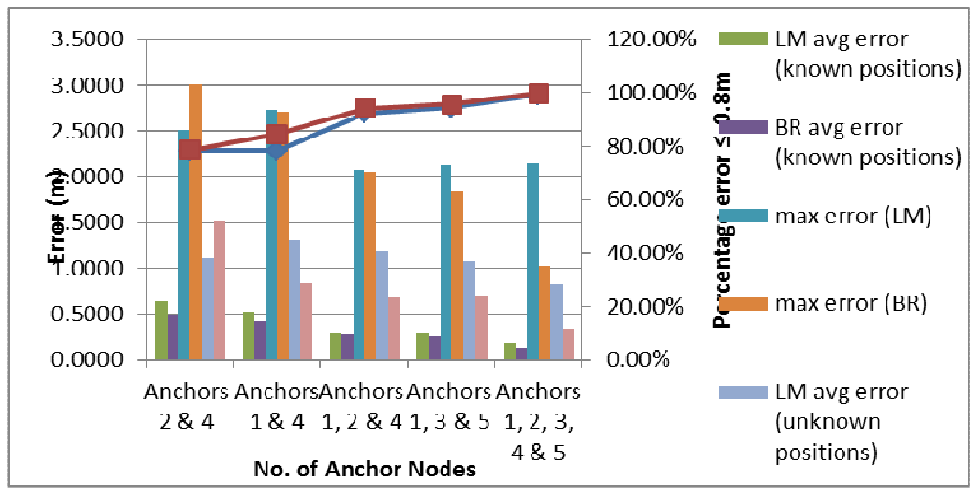

Figure 7. Localization error using LM and BR training algorithms for different anchor configurations.

From the results obtained it can be concluded that RSSI values can be used for localization where accuracy of less than a few meters is required. However, for applications requiring precise localization accuracy for indoor localization, additional sensors such as ultrasonic sensors [36] can be employed.

## 4.1. COMPARISON WITH OTHER RELATED LITERATURE

Various localization algorithms have been reported over the years with the goal of achieving higher localization accuracy at a low computational and hardware cost. Several concepts such as lateration, angulation, trilateration [25], multilateration and triangulation are used in localization. The various localization schemes can be categorised in one of the following groups: GPS based or GPS free, anchor based or anchor free, range based [7] or range free [2], [5]-[6], [23], fine grained or coarse grained, and stationary or mobile sensor nodes [22]. Several range based algorithms that have been proposed by a number of researchers include Time-Of-Arrival (TOA), Time Difference of Arrival (TDOA) [34], Angle Of Arrival (AOA) [9], [11] and Received Signal Strength Indicator (RSSI) [35]. As mentioned in Section 1, GPS localization schemes require external hardware, are costly and consume lot of energy, which is not suitable for WSNs. Use of directional antennas [21] have also been explored for the problem of node localization, with one or more nodes being used to measure the angle of arrival of the signal. The angle obtained is then used to compute the position of the node whose position is unknown. However, it must be noted that directional antennas are costly and require larger amount of power that leads to a WSN localization system which is not energy efficient.

The authors in [17] proposed a 2D and 3D localization algorithm using weighted centroid (WCL) technique. The number of anchors involved in this method is controlled by use of an optimized threshold. A range free WCL for 3D WSN using Mamdani & Sugano Fuzzy Inference System (FIS) is presented in [20]. The method involves computing the edge weights from the RSSI values by making use of the Mamdani & Sugano inference system where 121 anchor node were employed. The high-resolution range-independent localization (HiRLoc) [16] and modified HiRLoc [33] scheme used omni directional antennas. The average localization error, number of anchor nodes used and the implementation environment for some of the localization algorithms are given in table 1.

Table 1. Comparison of different localization algorithms.

| Localization Algorithm | Average Localization Error (m) | Number of anchor nodes | Implementation Environment |
|---|---|---|---|
| 2D-WCL [17] | > 3.000 | 100 | Simulation |
| Mamdani and Sugano FIS [20] | 3.0000 | 121 | Simulation |
| HiRLoc Scheme [16] | 3.6000 | | Simulation |
| Modified HiRLoc Scheme [33] | 3.5100 | | Simulation |
| Sequence-Based Localization [14] | 5.0000 | 10 | Simulation |
| Neural Network (3D) [18] | 0.4855 | 4 | Simulation |
| Proposed Method | 0.6887 | 3 | Real time |
| | 0.2953 | 4 | |
| | 0.3435 | 5 | |

In [18], a neural network approach for 3D localization of nodes using 4 anchor nodes is presented. Average localization error of 0.4855 m for 2D movement was reported. For 95% of the mobile nodes, a localization error of less than 0.8 m was achieved. A 10-10-3 neural network structure having four inputs was employed. The first layer used a hyperbolic tangent sigmoid activation function; the second layer used a log sigmoid activation function while the output used a liner activation function. However, for this research an average localization error of 0.2953 m was obtained using neural network with 4 anchor nodes, which is comparably lower than that of [18]. On the other hand, the authors in [18] presented their result based on simulation whereas in this research the actual implementation and testing has been carried out in real time environment.

## 5. CONCLUSIONS

An efficient 2D localization algorithm for WSN has been proposed by utilizing ANN. Different anchor node configurations have been evaluated that can be used by researchers in selecting the number of anchor nodes to use depending on their application environment. The input to the proposed method is the RSSI values of the signal received from the anchor nodes. As all sensor nodes are equipped with RF modules for wireless communication, no external hardware is required. The proposed method is thus energy efficient and uses only a two way message to obtain the inputs for the localization. However, it is recommended that the RSSI value of the localization beacon request signals received by the anchor nodes also be utilized that can result in a further reduction of the localization error.

For obtaining the best neural network for localization, the BR training algorithm is evaluated to give the best result. However, the training time for this algorithm is quite high compare to other methods such as LM, RP and SCG. Therefore, it is recommended to use the BR method when offline training is performed as in this case. For applications requiring online training, the LM method is recommended.

### ACKNOWLEDGEMENTS

This research work was fully supported by the College Research Committee of Fiji National University.

**AUTHORS**

**Shiu Kumar** received Bachelor of Engineering Technology and Postgraduate Diploma in Electrical and Electronics Engineering from the University of the South Pacific (USP) in year 2009 and 2012 respectively. He received his Masters in Electronics Engineering from Mokpo National University, South Korea. Currently he is pursuing his PhD in Machine Learning (Signal Processing) from USP. His research interests include Automation and Control, Wireless Sensor Networks, Embedded Microprocessor Applications, Artificial Intelligence and Signal Processing. He is a member of IEEE, IAENG and IACSIT.

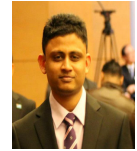

**Ronesh Sharma** received the BTech degree from the University of the South Pacific (USP), Suva, Fiji, in 2007 and MEng degree from Mokpo National University, South Korea. He is now pursuing his PhD degree in Engineering at University of the South Pacific, Suva, Fiji. His research interests include Bioinformatics, protein secondary, fold and structural class prediction problems, protein subcellular localization prediction problems, intrinsically disordered protein related problems, data mining, and pattern recognition. He is a member of IEEE.

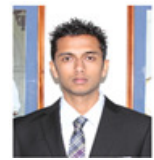

**Edwin Vans** was born in Ba, Fiji, on October 26, 1987. He received his Bachelor of Engineering Technology degree, Postgraduate Diploma in electrical and electronics engineering and MSc in engineering degree from the University of the South Pacific, Fiji in 2009, 2011 and 2015 respectively. He is currently a Lecturer in electronics engineering at School of Electrical and Electronic Engineering in Fiji National University. He is a current member of IEEE and IEEE Robotics and Automation Society. His research interests are in machine learning, intelligent robotics, and computer vision.

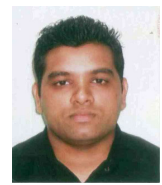